\documentclass[twoside,twocolumn,american,preprintnumbers,nofootinbib,aps,pra,10pt]{revtex4-1}

\usepackage{amsmath}
\usepackage{amssymb}
\usepackage{graphicx}
\usepackage{babel}
\usepackage{color}
\usepackage{hyperref}

\def\be{\begin{equation}}
\def\ee{\end{equation}}
\def\bea{\begin{eqnarray}}
\def\eea{\end{eqnarray}}

\begin{document}

\title{Non-Equilibrium Random Matrix Theory : Transition Probabilities}
\author{Francisco Gil Pedro}
\affiliation{Departamento de F\'{\i}sica Te\'orica and Instituto de F\'{\i}sica Te\'orica UAM/CSIC,Universidad Aut\'onoma de Madrid, Cantoblanco, 28049 Madrid, Spain}

\author{Alexander Westphal}

\affiliation{Deutsches Elektronen-Synchrotron DESY, Theory Group, D-22603 Hamburg, Germany}

\begin{abstract}
In this letter we present an analytic method for calculating the transition probability between two random Gaussian matrices with given eigenvalue spectra in the context of Dyson Brownian motion. We show that in the Coulomb gas language, in large $N$ limit, memory of the initial state is preserved in the form of a universal linear potential acting on the eigenvalues. We compute the likelihood of any given transition as a function of time, showing that as memory of the initial state is lost, transition probabilities converge to those of the static ensemble.
\end{abstract}

\preprint{DESY-16-104}
\preprint{IFT-UAM/CSIC-16-053}

\maketitle

{\bf I. Introduction}\;\; Since its inception, Random Matrix Theory (RMT) has proved a useful tool in the study of a variety of physical systems: from quantum mechanics to nuclear physics, from condensed matter to the string landscape, to name only a few. Beyond physics RMT has been used to describe the behaviour of correlation matrices wherever datasets at large-$N$ appear (including in recent years in areas like image analysis, genomics, epidemiology, engineering,
economics and finance, for reviews see e.g.~\cite{ANU:298726,Bouchard:2009qf}). The vast scope of these applications stems from the emergent simplicity of systems with many degrees of freedom and non-trivial interactions: it is  often the case that the generic properties of such complex systems are independent of the finer details, being instead determined by simpler structures. In the past RMT techniques developed in the study of a particular problem have provided valuable insights in the study of completely distinct physical systems. It is with this in mind that we present in this letter a new result that originated by applying Dyson Brownian motion (DBM) to Gaussian ensembles in the context of the string landscape and cosmic inflation \cite{Pedro:2016}, but that is general in RMT.

Some of the most common and useful applications of RMT are related to properties of the eigenvalue spectrum of a given matrix ensemble. Properties like the distribution of spacings between adjacent eigenvalues or the probability of occurrence of certain eigenvalue spectra are widely studied in the literature. Such questions can often be addressed by studying the appropriate probability density function (pdf), specially in the case of classical ensembles like the Gaussian/Wigner or Wishart where such a pdf is known. 

A crucial intuitive insight in the study of ensembles for which the pdf is known was provided by Dyson, who first related the pdf for a given matrix ensemble with the partition function for a gas of charged particles moving on the real line, subject to forces derived from the pdf. This interpretation is often called the Coulomb gas picture, and in it each eigenvalue configuration corresponds to a distinct thermodynamical equilibrium state to which one can assign an occurrence probability according to the pdf. In this framework there is neither a notion of time nor evolution between different states. Acquiring an analytic description of time-dependence of e.g. the eigenvalue density and related quantities under the process of DBM from non-equilibrium initial conditions is highly desirable in view of their importance for many large-$N$ random systems beyond theoretical physics.


{\bf II. Dyson Brownian Motion} Time dependence can be introduced in RMT by postulating that the matrices are undergoing Brownian motion according to  $M_{ij}(s+\delta s)=M_{ij}(s)+\delta M_{ij}$, where the variations have the following statistical properties \cite{Dyson:1962}
\bea
\langle\delta M_{ij}\rangle=-M_{ij}\frac{\delta s}{\sigma^2 f}\label{eq:drift}\ , \\
\langle(\delta M_{ij})^2\rangle= (1+\delta_{ij}) \frac{\delta s}{\beta f}\label{eq:diff}\ .
\eea
Here $s$ plays the role of time (though it is not necessarily related to physical time and its meaning will depend on the problem in question), $\sigma^2$ is the variance of the distribution from which the entries of $M$ are drawn, which we choose to be $\sigma^2=a/N$,  $f$ is a friction coefficient and $\beta$ is the ``temperature'' of the gas: $\beta=\{1,2,4\}$ for the Gaussian orthogonal (GOE), unitary (GUE) and symplectic (GSE) ensembles respectively.

The transition probability between an initial state matrix $M_0$ and a final state $M$ in a Wigner ensemble can be found by integrating the pdf \cite{Uhlenbeck:1930zz}
\be
d P=\mathcal{C} \ \exp \left\{-\frac{\beta}{2 \sigma^2(1-q^2)}\mathrm{Tr}[(M-q M_0)^2] \right\} dM_{ij}\ ,
\label{eq:jointpdf}
\ee
which is a solution to a Fokker-Planck equation describing the matrix Brownian motion. In \eqref{eq:jointpdf} $q$ is a reparametrization of the time coordinate $s$, $q\equiv\exp(-\frac{s}{\sigma^2 f})$ and $\mathcal{C}$ is an overall normalisation constant, ensuring that the probabilities are always smaller than unity. The aim of this letter is to present an analytic method to estimate the transition probability, in a generalisation of the time-independent method of \cite{Dean:2006wk,Dean:2008}.

{\bf III. Path integral representation of DBM}\;\; In the Coulomb gas picture, from the pdf of Eq. \eqref{eq:jointpdf} one can derive the Hamiltonian
\be
\begin{split}
H(\lambda)=&-\frac{\beta}{2}\sum_{i=1}^{N} \left\{ \frac{\lambda_i^2}{\sigma^2 (1-q^2)}-\frac{2 q \lambda_i M_0^{ii}}{\sigma^2 (1-q^2)}\right\}+\\
&+\frac{\beta}{2}\sum_{i\neq j} \ln|\lambda_i-\lambda_j |\ ,
\label{eq:Hlambda}
\end{split}
\ee
from which one reads the forces acting on the charged particles. The transition probabilities in the Brownian motion model are therefore determined by the interplay between a quadratic self-interaction, a logarithmic mutual repulsion (both present also in the static ensemble albeit here they have $q$-dependent magnitudes) and a linear potential. This last contribution describes the way the system preserves the memory of its initial configuration $M_0$, since the strength of the linear potential acting on the $i$-th eigenvalue $\lambda_i$ is set by the corresponding diagonal entry of $M_0$.

In order to make progress one must reduce the number of variables in the problem by approximating the different linear potentials acting on the $\lambda_i$ by a single universal potential: $M^0_{ii}= m, \ \forall \ i=[1,N]$. It seems natural to choose $m$ to be the mean diagonal entry of $M_0$, which is given by the mean eigenvalue of $M_0$ since
\be
m=\frac{1}{N}\sum_{i=1}^{N} M^0_{ii}= \frac{1}{N}\text{tr}[ M_0]= \langle \lambda_{M_0} \rangle \ .
\ee
This universal potential approximation is necessary in order to find a simple analytical estimate for the transition probability, and we'll see it captures the behaviour of the system away from the immediate vicinity of the initial point.  Furthermore we note that there is a one-to-one correspondence between the smallest eigenvalue of $M_0$ and its mean eigenvalue $\langle\lambda_{M_0}\rangle$, so even though some information about the initial state is lost, by specifying the mean force, in the large $N$ limit, one uniquely identifies the initial eigenvalue spectrum.

With this simplification in place one can proceed along the lines of \cite{Dean:2006wk,Dean:2008} and write the Hamiltonian of Eq. \eqref{eq:Hlambda} in terms of the empirical eigenvalue density function $\rho(\lambda)$ 
\be
\rho(\lambda)=\frac{1}{N}\sum_{i=1}^{N} \delta(\lambda-\lambda_i)\ ,
\ee
determine $\rho(\lambda)$ that minimises the "energy" of the system and subsequently approximate the integral by the maximum of the integrand.

Pulling out the overall constant factors and introducing a Lagrange multiplier $\alpha$ to ensure the correct normalisation of the density function one defines
\bea
\Sigma[\rho]&=&\frac{1}{2 \tilde{a}}\int\lambda^2\  \rho(\lambda)\ d\lambda+\frac{b}{2 \tilde{a}}\int\lambda\ \rho(\lambda)\ d\lambda
\\&&\hspace{-0.5cm}-\frac{1}{2}\int\hspace{-0.1cm} \ln|\lambda-\tilde{\lambda}|\rho(\lambda)\rho(\tilde{\lambda})\ d\lambda\ d\tilde{\lambda}+\alpha\left\{ \int\hspace{-0.1cm} d\lambda\ \rho(\lambda)-1\right\}\ \nonumber
\label{eq:E}
\eea
to leading order in $N$. In Eq. \eqref{eq:E} we have introduced the shorthand notation $\tilde{a}\equiv a (1-q^2)$, $b\equiv -2 qm$. 

{\bf IV. Saddle point evaluation}\;\; Transition probabilities are now computed as functional integrals over the space of suitably normalised eigenvalue density functions
\be
P\propto \int \exp \left \{-\beta N^2 \Sigma[\rho] \right \} \ d[\rho]\  d\alpha
\ee
and are dominated by the eigenvalue configuration $\rho_c(\lambda)$ that minimises the energy $\Sigma$ of the system:
\be
\frac{d \Sigma}{d \rho}\Big |_{\rho_c}=0 \Leftrightarrow \frac{\lambda^2}{2 \tilde{a}}+\frac{b \lambda}{2 \tilde{a}}-\int d\tilde{\lambda}\ \rho_c(\tilde{\lambda})\  \log|\lambda-\tilde{\lambda} |+\alpha=0.
\label{eq:9}
\ee
Differentiating Eq. \eqref{eq:9} w.r.t. $\lambda$ one finds the following integral equation for $\rho_c(\lambda)$ :
\be 
\frac{\lambda}{\tilde{a}}+\frac{b}{2\tilde{a}}=\mathrm{P}\int_{\zeta}^\infty d\tilde{\lambda} \ \frac{\rho_c(\tilde{\lambda})}{\lambda-\tilde{\lambda}}\ ,
\label{eq:tricomi0}
\ee
where $P$ stands for the Cauchy's principal value of the otherwise ill-defined integral. Solving integral equations of the type of \eqref{eq:tricomi0} is in general non-trivial, however in this case, a simple shift of the integration variables ($\tilde{x}\equiv\tilde{\lambda} -\zeta$ , $x\equiv\lambda -\zeta$) recasts Eq. \eqref{eq:tricomi0} in a form where one can use Tricomi's theorem \cite{tricomi} to find the time dependent eigenvalue density function
\be\label{eq:eigvaldensity}
\rho_c(x)=\frac{1}{2 \pi \tilde{a}}\sqrt{\frac{L-x}{x}}[L+2(x+\zeta+b/2)]\ ,\;\; x\in [0,L]
\ee
that minimises the "energy" of the eigenvalue system at a given time $q$. Note that $\rho_c(x)$ vanishes outside the interval. The parameter $L$ defines the domain of $\rho_c$ and is determined by imposing the proper normalisation of the density function $\int_0^L \rho_c \ dx =1$ :
\be
L=\frac{2}{3}\left[-(\zeta+b/2)+\sqrt{6\tilde{a}+(\zeta+b/2)^2}\right] \ .
\ee

\begin{figure}[t]
\begin{center}
\includegraphics[width=0.45\textwidth]{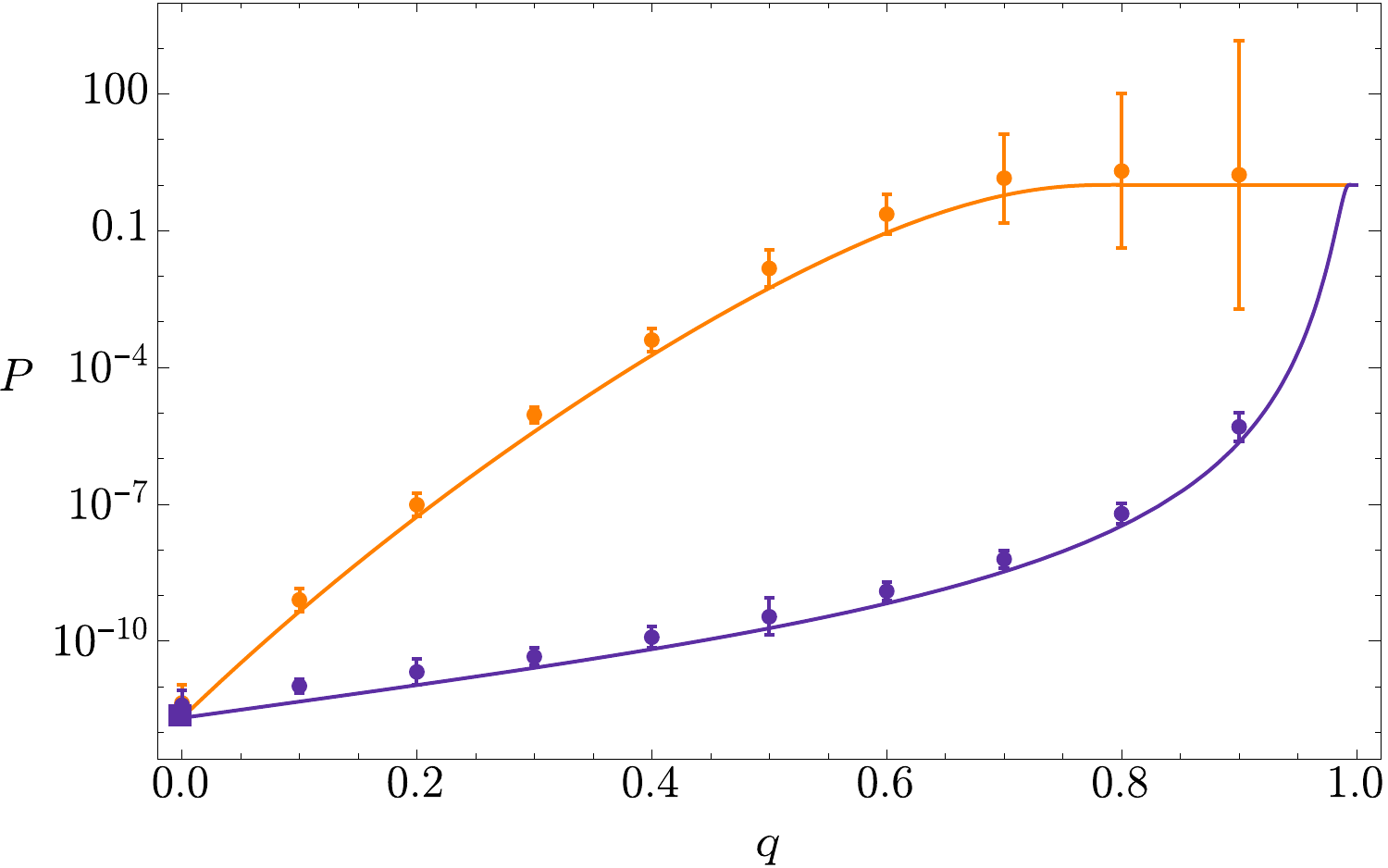}
\caption{Transition probabilities towards a positive-definite matrix from two distinct initial states. 
 Points with error bars correspond to numerical integration of the pdf, lines correspond to the analytical estimate using the rate function of Eq. \eqref{eq:Psi}.}
\label{fig:Ptrans}
\end{center}
\end{figure}

The minimal energy of the system is promptly found to be 
\bea
\Sigma[\rho_c]&=&\frac{\zeta^2+b\ \zeta}{4 \tilde{a}^2}+\frac{1}{4\tilde{a}}\int\lambda^2\ \rho_c(\lambda) \ d\lambda+\frac{b}{4 \tilde{a}} \int \rho_c(\lambda)\ \lambda\ d\lambda\nonumber\\
&&-\frac{1}{2}\int d\tilde{\lambda} \ln |\zeta-\tilde{\lambda}|\ \rho_c  (\tilde{\lambda})\ ,
\label{eq:Sig}
\eea
where the integrals are to be performed over the interval $[\zeta,L+\zeta]$. 
Since $\rho_c$ is a density function it must be positive definite over its domain, in particular one must have $\rho_c(x=0)>0$, which implies that the Eq. \eqref{eq:Sig} is valid as long as 
\be
\zeta>-\sqrt{2 \tilde{a}}-\frac{b}{2}\equiv \zeta_{edge}.
\ee
Should the above bound be violated, the energy of the system is instead given by
\be
\Sigma[\zeta<\zeta_{edge}]=\Sigma[\zeta_{edge}].
\ee
Note that in the case of the static ensemble ($q=0$, $b=0$, $\tilde{a}=a$) this corresponds to looking for matrices with $\forall \ \lambda > \zeta > \zeta_{edge}=-\sqrt{2 a }$, i.e. with all eigenvalues larger than the left edge of the Wigner semi-circle. These configurations have the same energy as the Wigner semi-circle.

{\bf V. Time-dependent transition probability - the rate function}\;\; To leading order in $N$ the transition probability is given by 
\be
P(M(s),M_0)=\exp\left\{-\beta N^2(\Sigma[\zeta]-\Sigma[\zeta_{edge}])\right\} \ ,
\label{eq:PP}
\ee
where the subtraction of $\Sigma[\zeta_{edge}]$ is necessary for the correct normalisation of the transition probability. Equation \eqref{eq:PP} prompts us to define the rate function $\Psi(\tilde{\zeta})\equiv\Sigma[\zeta]-\Sigma[\zeta_{edge}]$ which in light of Eq. \eqref{eq:Sig} can be written as 
\bea
\Psi(\tilde{\zeta})&=&\frac{1}{108  \tilde{a}^2}\left\{ 36 \tilde{a} \tilde{\zeta}^2-\tilde{\zeta}^4+(15 \tilde{a} \tilde{\zeta}+\tilde{\zeta}^3)\sqrt{6 \tilde{a}+\tilde{\zeta}^2} +\right .\nonumber\\
&&\left.\hspace{-0. cm}+ 27 \tilde{a}^2\left[\ln(72\tilde{a}) -2\ln(2(\sqrt{6 \tilde{a}-\tilde{\zeta}}-\tilde{\zeta}))\right]\right\}
\label{eq:Psi}
\eea
with $\tilde{\zeta}\equiv \zeta+b/2$, such that the transition probabilities are given, to leading order, as 
\be
P(M(s),M_0)= \exp{\left[-\beta N^2 \Psi(\tilde{\zeta})+\mathcal{O}(N)\right]}\ .
\label{eq:Pfinal}
\ee

In Fig. \ref{fig:Ptrans} we plot the evolution of the probability corresponding of transitions to a positive definite matrix ($M \ : \  \forall \ \lambda >0 $)  from two different initial states:  $M_0^A \ : \  \forall \ \lambda >1 $ and $M_0^B \ : \  \forall \ \lambda >-1 $. Points with error bars correspond to data obtained by numerical integration of Eq. \eqref{eq:jointpdf}, while lines denote the analytical result of Eqs. \eqref{eq:Psi} and \eqref{eq:Pfinal}, both assuming a  universal linear potential. The transition $M_0^A\rightarrow M$ corresponds to relaxation of the system towards a configuration with less "energy" and therefore one sees that in the first correlation length ($0.37<q<1$) it is very likely to take place, $P\sim 1$, after which point it decays exponentially with $q$. The case $M_0^B\rightarrow M$ corresponds to a transition in the direction opposite to the natural relaxation of the system, and therefore one expects the corresponding probability to decrease very rapidly with $q$. In both cases the end point of the evolution at future infinity, $q=0$, when all memory of the initial configuration has been lost, is given by the probability of drawing the matrix $M$ from the static ensemble. This asymptotic result at $q\to 0$ was first estimated in \cite{Dean:2006wk,Dean:2008} by a saddle point computation analogous to the one performed here. We conclude that, under the approximation of a universal linear potential, the saddle point method constitutes a good approximation to the integration of the pdf. It remains to be seen how this approximation fares when compared against numerically generated DBM data, following from Eqs. \eqref{eq:drift} and \eqref{eq:diff}.   

\begin{figure}[t]
\begin{center}
\includegraphics[width=0.48\textwidth]{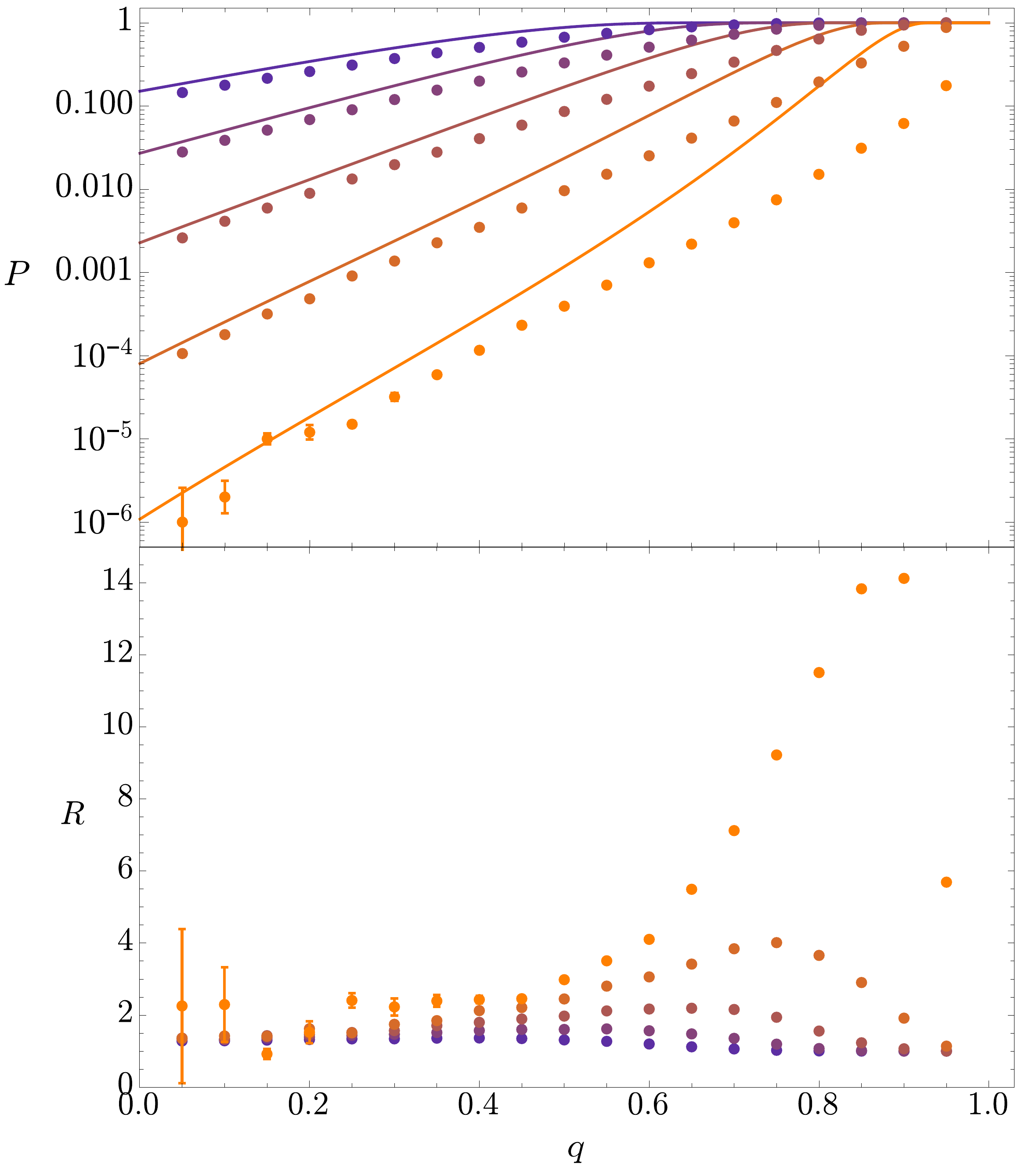}
\caption{Transition probability estimates (see main text for details).
 The dashed lines corresponds to the saddle point approximation to the pdf integration with a universal linear potential; points correspond to matrix DBM according to Eqs. \eqref{eq:drift}-\eqref{eq:diff}. The top panel shows the $q-$dependence of the transition probability, while the bottom shows the evolution of the ratio between the analytical estimate and the actual DBM behaviour.}
\label{fig:Ptrans2}
\end{center}
\end{figure}

In Fig.~\ref{fig:Ptrans2} we evaluate the validity of the universal linear potential approximation by comparing the analytical estimate (lines) with numerical DBM data (points). We choose initial spectra $M_0 \ : \  \forall \ \lambda  > 0 $  and look for fluctuations to spectra $M(s)\ : \  \forall \ \lambda >\zeta $, with $\zeta=\{ -1,-0.75,-0.5,-0.25,0 \}$ in the GUE ($\beta=2$) for $N=5$ \footnote{To remove ambiguities in the choice of initial conditions, $M_0$ is set to be the matrix whose eigenvalues are the average eigenvalues of a large set of matrices satisfying the $\forall \ \lambda  > 0$ cut. }. We see that the analytical estimates always fall within the order of magnitude of the numerical results after one correlation length $q<e^{-1}\sim0.37$. Note that in the regime $0<q<e^{-1}$ the transition probability is varying significantly and that in this range the precision of our approximation is at the level of best analytical methods available for the static ensemble (which are captured in the $q\rightarrow0$ limit Fig. \ref{fig:Ptrans2}) \cite{Dean:2006wk,Dean:2008}. 

In fact the results of the saddle point computation, the time-dependent fluctuated eigenvalue density Eq.~\eqref{eq:eigvaldensity} and the time-dependent rate function Eq.~\eqref{eq:Psi}, allow us now to provide a more detailed analytical justification of the single linear-potential approximation and the universal $\Psi(q)=\Psi(0)-{\cal O}(q)\,,\,q\lesssim e^{-1}$ behaviour of the rate function. For this purpose, we look at the regime where the initial conditions are provided by spectra with eigenvalues larger $\zeta_i=-2+\epsilon$, while selecting for final spectra with all eigenvalues larger than $\zeta_f=-2+\delta$. We can now expand the rate function in small $\epsilon,\delta$ at small $q$, finding $\Psi=\frac{\delta^3}{12}\left(1+\frac32 q^2\right)-\frac{1}{16} \,\delta^2\epsilon^2 q $ to leading order. The $q$-independent piece is compatible with the result of Dean and Majumdar for the static ensemble. We see that the $q$-dependent dominant piece at small $q$ is linear in $q$, and its coefficient is uniquely determined by the edges of the fluctuated initial and final condition choices. For small enough $q$ the rate function always becomes linear in $q$ which explains the universal behaviour at late times. We see moreover, that $\Psi$ becomes linear in $q$ typically within one correlation length, unless we ask for the final conditions given by spectra more unlikely than the initial conditions, that is $\delta > \epsilon$. In this particular regime the $q^2$ term in $\Psi$ will give way to the linear term only at successively smaller values of $q$. We can see this to begin happening in both the analytic behaviour of $\Psi$ and the DBM result in Fig.~\ref{fig:Ptrans2} around values of $q\simeq e^{-1}$ for the lowest set of curves where both initial and final conditions are given by all eigenvalues larger than $\zeta_i=\zeta_f=0$.

Let us now consider the relaxation process of DBM. It is clear that, whatever the initial conditions, relaxation drives the eigenvalue spectra to approach their static equilibrium configuration for $q\to 0$ at a rate estimated in \cite{Marsh:2013qca}. Now assume that we stop the relaxation process momentarily at some small value $q$. The  eigenvalue spectrum of a large set of matrices there is already close to the static configuration. Hence, at a given small $q$ we will find, with probability close to unity, only eigenvalue spectra of slightly fluctuated Wigner ensembles with a lower edge close to the semi-circle value $\zeta(q)=-2+\epsilon(q)$, where $\epsilon(q)$ depends on the spectrum at $q=1$. These slightly fluctuated Wigner ensembles thus form the DBM-produced most probable \emph{local} initial conditions for relaxing towards further decreasing $q$. 

Next, we let the relaxation process resume and focus on the simple $N=2$ case in order to extract analytical results. The resulting two linear potentials, and the average linear potential acting on the slightly fluctuated Wigner ensemble of matrices $M_q$ describing the local initial conditions at the given $q\ll 1$ are now given in terms of a slightly fluctuated ensemble with edge at $\zeta(q)=-2+\epsilon(q)$. 

Therefore, to get the two linear potentials and their variances acting at that given $q\ll1$, we now compute $\langle\lambda_{M_q}^i\rangle\equiv \langle M_q^{ii}\rangle$ governing their strength, and their respective variances $\langle(M_q^{ii}-\langle\lambda_{M_q}^i\rangle)^2\rangle$. These quantities provide the estimators of the two linear potentials and their variances, respectively. 
In general, when passing to eigenvalues densities, the Hamiltonian contains $N$ linear potentials given by the expectation values of $M_q^{ii}$. For $N=2$ we can estimate the strength of these forces by estimating $\langle M_q^{ii}\rangle$ from the static fluctuated eigenvalue density $\rho_c(\mu, q=0)$ of Eq.~\eqref{eq:eigvaldensity} as follows
\be\begin{split}
\langle \lambda_{M_q}^1\rangle &= \int \limits_{-2+\epsilon}^{\epsilon^2/8}\hspace{-1ex}d\mu\Big(\mu\,\rho_c(\mu,q=0)\Big)+{\cal O}(q)\\
\label{eq:Nflinearforces}
\langle \lambda_{M_q}^2\rangle &= \int \limits_{\epsilon^2/8}^{2+\epsilon^2/16}\hspace{-3ex}d\mu\Big(\mu\,\rho_c(\mu,q=0)\Big)+{\cal O}(q)\quad.
\end{split}
\ee
Here, the upper integration boundaries are determined to leading order in the lower edge shift $\epsilon$ by demanding equal probability for the left and right half-bands of the eigenvalue density  $\int_{-2+\epsilon}^{\epsilon^2/8}d\mu\rho_c(\mu,q=0)=\int_{\epsilon^2/8}^{2+\epsilon^2/16}d\mu\rho_c(\mu,q=0)=1/2+{\cal O}(\epsilon^3)$. We then find that $\langle \lambda_{M_0}^{1/2}\rangle=\pm  \frac{8}{3\pi}+{\cal O}(\epsilon^2)$ such that the total average eigenvalue $\langle\lambda_{M_0}\rangle={\cal O}(\epsilon^2)$, while the variances of both $\langle\lambda_{M_0}\rangle$ and $\langle\lambda_{M_0}^i\rangle$ deviate from their unfluctuated semi-circle values only at ${\cal O}(\epsilon^2)$ as well. Note, that this result is consistent with the fact that the ${\cal O}(q)$-term in the rate function describing the dominant linear potential term  at  small $q$ is ${\cal O}(\epsilon^2)$. 

Hence, we find that the estimators for the eigenvalues giving the two linear potentials, $\langle\lambda_{M_0}^i\rangle$, as well as their variances deviate from the values for the unfluctuated semi-circle at an order \emph{suppressed} in $\epsilon\ll 1$ compared to the deviation of the edge of the intermediate initial condition spectrum at $q\ll1$ which is $\zeta(q)-\zeta(semi-circle)=\zeta(q)+2={\cal O}(\epsilon)$. For the exact semicircle the mean of the two equal-size linear potentials vanishes. Moreover, since for the semicircle the total linear potential vanishes, the effect of their equal-size variances of the semicircle distribution on the effective average linear potential must cancel out as well. Therefore, for spectra close to the semicircle, the averaged linear potential can only depend on the \emph{shift} of the $N$ linear potentials and their variances away from their  semicircle values. This shows that, at small $q$, \emph{the effects of having two linear potentials are given by just the single overall linear potential} given by $\langle\lambda_{M_0}\rangle=1/2 \sum_{i=1,2}\langle\lambda_{M_0}^i\rangle$ up to and including the second moments of the individual linear potentials at ${\cal O}(\epsilon^2)$. Hence, at small enough $q$ the single linear potential approximation becomes a good description, which a posteriori justifies the use of this approximation.

{\bf VI. Discussion}\;\; In this letter we have proposed an analytical extension of the description of non-equilibrium RMT through time-dependent Brownian motion.
We were able to use and extend the path integral representation and saddle point methods given in~\cite{Dean:2006wk,Dean:2008} to analytically derive the time-dependent eigenvalue density and transition probability rate function for a perturbed non-equilibrium Gaussian random system described by the fluctuated Wigner ensemble. Our results are general and hold for all Wigner ensembles $\beta=1,2,4$, which should give them wide applicability wherever RMT holds sway. 
We conclude by pointing out an example of an application of the method to cosmological inflation occurring in a string landscape modelled by a Gaussian ensemble. Successful inflation involves both having both a suitable critical point \emph{and rolling into a viable local minimum after inflation - the graceful exit}. It is exactly the probability of achieving such a graceful exit which our treatment of transition probabilities in DBM allows us to compute. Based on our previous work~\cite{Pedro:2013nda}, this enables us to derive analytical expression for the joint probability of inflation occurring \emph{together a graceful exit via rolling into a viable post-inflationary minimum} in the forthcoming work~\cite{Pedro:2016}.

\vspace{0.5cm}
\paragraph*{\it Acknowledgments:}
FGP's work is supported by the grants  FPA2012-32828 from the MINECO, the ERC Advanced Grant SPLE under contract ERC-2012-ADG-20120216-320421 and the grant SEV-2012-0249 of the ``Centro de Excelencia Severo Ochoa" Programme.  The work of A.W. is supported by the ERC Consolidator Grant STRINGFLATION under the HORIZON 2020 contract no. 647995.

\end{document}